\documentclass[manuscript,screen,nonacm]{acmart}
\AtBeginDocument{%
  \providecommand\BibTeX{{%
    \normalfont B\kern-0.5em{\scshape i\kern-0.25em b}\kern-0.8em\TeX}}}

\setcopyright{acmcopyright}
\copyrightyear{2021}
\acmYear{2021}

\usepackage{diagbox}
\usepackage{array}
\usepackage{tabularx}
\usepackage{graphicx}
\usepackage{subcaption}
\usepackage{soul}
\usepackage{devanagari}
\begin{document}


\title{From Assistants to Friends: Investigating Emotional Intelligence of IPAs in Hindi and English}

\author{Mallika Subramanian}
\authornote{Both authors contributed equally to this research.}
\email{mallika.subramanian@students.iiit.ac.in}
\author{Shradha Sehgal}
\authornotemark[1]
\email{shradha.sehgal@students.iiit.ac.in}
\author{Nimmi Rangaswamy}
\email{nimmi.rangaswamy@students.iiit.ac.in}
\affiliation{%
  \institution{International Institute of Information Technology, Hyderabad}
  \state{Telangana}
  \country{India}
}







\renewcommand{\shortauthors}{}

\begin{abstract}
Intelligent Personal Assistants (IPAs) like Amazon Alexa, Apple Siri, and Google Assistant are increasingly becoming a part of our everyday. As IPAs become ubiquitous and their applications expand, users turn to them for not just routine tasks, but also intelligent conversations. In this study, we measure the \textbf{emotional intelligence (EI)} displayed by IPAs in the English and Hindi languages; to our knowledge this is a pioneering effort in probing the emotional intelligence of IPAs in Indian languages. We pose utterances that convey the `Sadness' or `Humor' emotion and evaluate IPA responses. We build on previous research to propose a quantitative and qualitative evaluation scheme encompassing new criteria from social science perspectives (display of empathy, wit, understanding) and IPA-specific features (voice modulation, search redirects).

We find EI displayed by Google Assistant in Hindi is comparable to EI displayed in English, with the assistant employing both voice modulation and emojis in text. However, we do find that IPAs are unable to understand and respond intelligently to all queries, sometimes even  offering counter-productive and problematic responses. Our experiment offers evidence and directions to augment the potential for EI in IPAs.

\end{abstract}

\ccsdesc[500]{Human-centered computing~HCI design and evaluation methods}
\ccsdesc[500]{Human-centered computing~Ubiquitous and mobile computing design and evaluation methods}
\ccsdesc[300]{Human-centered computing~Natural language interfaces}
\ccsdesc[100]{Human-centered computing~Ambient intelligence}

\keywords{Virtual Assistants, IPAs, Emotional Intelligence, Human-Computer Interaction, Human-Machine Interaction}



\maketitle


\section{Introduction}
An Intelligent Personal Assistant (IPA) is an artificially intelligent system that can perform tasks or answer queries based on commands and questions ~\cite{iva_def}. 
As IPAs gain popularity, users turn to them not only for regular tasks, but also to express emotion. Therefore, many IPAs are marketed as a user’s companion with a human-like personality, rather than just a task-doer. Whilst much research has gone into studying the abilities of an IPA to understand and perform tasks \cite{iva_survey}, the field of emotional intelligence displayed by IPAs remains a relatively under-explored research area.

\par
The study of IPAs becomes important today as their usage has seen a sharp increase during the COVID-19 pandemic\footnote{\url{https://voicebot.ai/2020/05/01/voice-assistants-see-uptick-in-daily-use-during-the-pandemic-report/}}. Apart from being a hands-free technology, users are increasingly using IPAs as a “frustration outlet” \footnote{\url{https://www.washingtonpost.com/technology/2020/07/01/alexa-siri-google-home-assistant}}. In this context, researching emotional intelligence displayed by IPAs acquires renewed relevance.

\par 
Emotional Intelligence for humans is defined as the ability to sense, understand, value and effectively apply the power of emotions as a source of human energy, information, trust, creativity and influence ~\cite{ei_defn}. As machines cannot sense and feel emotion, we measure if an IPA can `apply the power of emotion' in their responses ~\cite{ei_ipa} i.e. whether the response entails a recognition of emotion and displays empathy, as a human's response would. For humans, EI is tested in 3 major forms - self reported, reported by others, and ability tests. As IPAs cannot self-report, we conduct ability tests based on the Mayers-Salovey-Calousey-Emotional Test (MSCEIT) \cite{msceit}. To evaluate EI in an IPA, we check for `portrayal' of emotion in the responses, rather than feeling of emotion. In this work, we explore the `Sadness' and `Humor' emotions in depth, in order to test 2 contrasting aspects of EI. In the `Sadness' category, we include prompts where the user has a sad tone. We test for empathy and understanding displayed by an IPA and check if it offers ways to uplift one's mood. On the other hand, the `Humor' category affords researchers to test a more `light-hearted' context and if indeed an IPA displays wit and can partake of a cheerful conversation.   

\par

We propose a nuanced evaluation scheme with quantitative features tailored to each category -- \textit{`Sadness’} and \textit{`Humor’} -- by extending previous approaches (Table \ref{table:metric_table}). To the best of our knowledge, ours is the first work to explore emotional intelligence displayed by IPAs in Indian languages. Many technology companies have announced their commitment to language diversity in IPAs \footnote{\url{https://blog.google/technology/next-billion-users/google-for-india-2018/}} with Apple announcing the support for 9 Indian languages on Siri, during WWDC 2021 \footnote{\url{https://support.apple.com/en-in/HT212537}}. As the support for Hindi in IPAs is relatively new, it is critical to study EI performance in the Hindi language and its standing with that of the English language. 



\par
We study EI along `Sadness' and `Humor' categories by posing a myriad of queries to different IPAs, in Hindi and English. We annotate 402 such question-response pairs along a new quantitative and qualitative metric. We make our dataset public for research. Our main contributions are:

\begin{enumerate}

    \item \textit{Comparison of EI relating to `Sadness' and `Humor' emotions displayed by IPAs (Siri, Google Assistant, and Alexa), along with key visualizations to understand and interpret results.}

    \item \textit{First evaluation of emotional intelligence in the Hindi language.}

    \item \textit{Proposal of a rigorous qualitative and quantitative evaluation scheme to test EI ability.}
    
\end{enumerate}

\section{Related work}

Studying EI in IPAs is an emerging research direction in Human-Computer Interaction. Previous studies have explored the applications of EI in IPAs. The role of IPAs in combating isolation and strengthening bonds among the elderly is one of them \cite{elderly_social_bond}.

Users of IPAs often attribute personality dimensions and anthropomorphic features to virtual assistants \cite{va_gender_pronouns, personification_alexa}. A user's perceptions about the humanness of an IPA is determined by the extent to which the agent is capable of showing meaningful emotions in their responses \cite{age_artificial_ei,role_of_emotion}. The introduction of human-like qualities like voice modulation induces a sense of anthropomorphism of machines and agents, which makes human users ascribe EI qualities -- like trust, empathy, and support -- to IPAs \cite{google_nbu_voice, autonomous_vehicles_mind}.  We harness the above to build our evaluation metric.

Exploring evaluation of EI in IPAs, a study of humorous interactions in English,  \cite{humour_classification} recruited participants to conduct week-long interactions with an IPA (one of Apple Siri, Amazon Alexa, or Google Assistant). The users then rated the humor level of IPA responses on a Likert-scale (1-5)  where over 50\%+ of the agent's utterances were rated as funny. In addition to using the notion and connotation of an IPA’s response in characterizing its EI in humorous scenarios, the underlying latent semantic structures such as ambiguity, interpersonal effect, phonetic style -- also played an important role \cite{latent-humour}. In order to analyze EI for mental and physical health related queries, a study \cite{jamainternmed} characterized the responses of IPAs based on various metrics, such as - 1. "recognition" -- if the IPA was able to identify the user query, 2. "respect" -- based on clinical experience with respectful language, and 3. "reference" -- whether or not the IPA refers to a helpline or contact number for the emergency situation.

Research has also suggested improvements to EI displayed by IPAs. For example, \cite{personality_perceived_ei} is a proposal to improve an IPA’s perceived EI through personality-driven expression of emotions to complete user tasks.

Our paper takes inspiration from Yang et. al.’s research on measuring EI in virtual assistants \cite{perceived_ei}. They propose an interactive dialog system (Zara) and compare a non-emotion expressing VA with one that expresses emotions exploring both verbal and visual aspects of communication. Participants of the study interacted with the IPAs and evaluated them via a questionnaire that was created based on the MSCEIT \cite{msceit} for scoring the IPAs. However, their work focuses on the English language while we build on their evaluation methodology to evaluate and deploy the \textbf{Hindi language}. We leverage studies \cite{msceit, perceived_ei} that utilize pre-defined metrics to quantify the performance of IPAs and combine this with additional features based on social science perspectives, custom to each category of emotion -- Eg: analyzing the \textit{wit} quotient, \textit{voice modulation} and \textit{use of references} in humour, \textit{supportive responses} in sadness -- and propose a new metric to quantitatively evaluate the EI performance of IPAs.

\section{Data}

We pose queries to test the emotional intelligence of 3 different IPAs -- Amazon Alexa, Apple Siri, and Google Assistant. We divide the queries into 2 categories -- `Sadness' and `Humor'. Queries in the `Sadness' field include statements where the user claims to be unhappy or shares a sorrowful event occurrence -- example queries include “I am lonely” and “I am feeling sad”. `Humor' related queries test an agent's quips and evaluate the funny quotient and wit displayed in their responses. The question categories include: Personality, eg: \textit{"What's your favourite colour?"}, Rhetoric, eg: \textit{"Where am I?"}, Joke, eg: \textit{"Why did the chicken cross the road?"} and Reference based, eg: \textit{"Do you want to build a snowman?"}. At the time of writing this paper, Hindi language was unavailable in Siri, so we tested it on Amazon Alexa and Google Assistant. Overall, we pose 402 queries in total, with 156 in Hindi and 246 in English. The exact set of questions posed in each category can be found \href{https://docs.google.com/spreadsheets/d/1bTHtQdF9-yqOmNldhvTdEgeadjff4_xy-EdRnLiSHMU/edit#gid=0}{here}.

    



\section{Evaluation Metric}


As per the Mayers-Salovey-Calousey-Emotional Test (MSCEIT) \cite{msceit}, emotional intelligence can be measured along four branches - \textit{perceiving, using, understanding}, and \textit{managing emotions}. Previous studies on IPA evaluation have used this basic framework to create more fine-grained features \cite{perceived_ei}. Extending their methodology, we combine existing categories with new \textbf{quantitative categories} for evaluation and analysis. 

Whilst investigating the `Sadness' emotion, we use the categories `Identification` (\textit{perceiving branch}) and `Empathy` (\textit{understanding branch}) defined by previous works \cite{perceived_ei, empathetic_chatbots_mental_health}. We add categories for if the response is \textbf{uplifting} and if the agent \textbf{offers help} in the \textit{managing emotions} branch of MSCEIT. Furthermore, we check if the agent gives an entirely opposite response for a sadness-related query -- such as "I am happy for you" or "That's awesome!" to a query like "I am good for nothing". For `Humour` as an emotion, we build on the work by \cite{humour_classification} by adding categories that check for the use of \textbf{voice modulation} \cite{acoustic_cues_voice, speech_voice_NUI, google_nbu_voice, latent-humour}, \textbf{recognition/use of references} (under the \textit{using emotion} branch) and \textbf{wit, sarcasm/irony} in the responses, under the \textit{understanding} branch. 


We also add \textbf{qualitative features} such as checks for variance in agent responses for repeatedly posed queries; use of emojis; and a categorical label for the `Humor' category queries as -- jokes, rhetorical, personality or reference based questions. We add IPA-specific features such as \textbf{search redirects} under the \textit{perceiving emotion}, to check the frequency of IPAs relying on the web to find a response. This often occurs when the agent is not able to make sense of the query and simply returns search results on the topic, thereby showing poor EI. Evaluation features across all branches -- \textit{perceiving, using, understanding} and \textit{managing} and categories -- `Sadness' and `Humor' can be found in Table \ref{table:metric_table}.

In our coding methodology, we mark each category (or feature) on a binary scale of 0 and 1 where 1 indicates the presence of a feature and 0 indicates its absence. Few examples are given in Table \ref{tab:coding_example_google}.


\begin{table}[ht]
    \centering
    \begin{tabular}{|>{\raggedright\arraybackslash}p{0.15\linewidth}|>{\raggedright\arraybackslash}p{0.15\linewidth}|>{\raggedright\arraybackslash}p{0.15\linewidth}|>{\raggedright\arraybackslash}p{0.15\linewidth}|>{\raggedright\arraybackslash}p{0.15\linewidth}|>{\raggedright\arraybackslash}p{0.15\linewidth}|}\hline
        \diagbox[width = 1.1\linewidth]{Emotion}{Branch} & \textbf{Perceiving} & \textbf{Using} & \textbf{Understanding} & \textbf{Managing} & \textbf{Miscellaneous}\\\hline\hline
        \textbf{Sadness} & Search Redirect, Identify sadness &  Opposite Response & Empathy displayed & Uplifting response / Help & Varies for multiple attempts\\\hline
        \textbf{Humour} & Search redirect, Recognize/Use Reference & Voice Modulation & Wit, Sarcasm/Irony & Funny Quotient & Emoji effects, Joke Type \\\hline
    \end{tabular}
    \caption{Metric definition for evaluating emotional intelligence of IPAs. Using the 4 branch test model from MSCEIT \cite{msceit} we categorize features under each of the branches in order to quantify our analysis of EI in intelligent personal assistants.}
    \label{table:metric_table}
\end{table}

\begin{table}[!ht]
    \centering
    \begin{tabular}{|>{\raggedright\arraybackslash}p{0.17\linewidth}|
    >{\raggedright\arraybackslash}p{0.17\linewidth}|
    >{\raggedright\arraybackslash}p{0.04\linewidth}|
    >{\raggedright\arraybackslash}p{0.1\linewidth}|
    >{\raggedright\arraybackslash}p{0.07\linewidth}|
    >{\raggedright\arraybackslash}p{0.15\linewidth}|
    >{\raggedright\arraybackslash}p{0.1\linewidth}|
    >{\raggedright\arraybackslash}p{0.05\linewidth}|}
    \hline
    Query & Response & \multicolumn{2}{c|}{Perceiving} & Using & Understanding & Managing & Misc\\
    \hline
    \multicolumn{2}{|c|}{} & ID & SD & OR & ED & UR & VM \\
    \hline
    Nobody loves me & I am here for you & 1 & 0 & 0 & 1 & 0 & 0 \\ \hline 
    I just got out of a relationship & I am so happy for you & 0 & 0 & 1 & 0 & 0 & 1 \\ 
    \hline
    \end{tabular}
    \caption{Binary scale coding examples for Google Assistant's responses to Sadness emotion related queries. ID: Identify, SD: Search Redirect, OR: Opposite Response, ED: Empathy Displayed, UR: Uplifting Response,  VM: Varies for Multiple  }
    \label{tab:coding_example_google}
\end{table}

\vspace{-2.5em}
\section{Annotations}

For annotating and evaluating the responses of the IPAs, we approached three native Hindi speakers, who are also fluent in English, ranging from ages 20-25 years. Annotators had prior experience with using all 3 IPAs for their routine tasks or academic experiments. The annotaters marked the quantitative features mentioned above on a 0-1 scale.
For the qualitative analysis, annotators could fill out the comments section against each entry, with remarks about the appropriateness or creativity of the IPA response, as a way to capture miscellaneous observations. The annotators were also told to highlight the problematic or counter-productive responses where the agents failed to display EI.


\section{Findings}
The IPAs' performances vary to different degrees across our metric's categorical branch attributes, especially in certain nuanced categories such as -- voice modulation, emoji usage, search redirects. 
We found that Google Assistant's responses displayed similar emotional intelligence for both English and Hindi queries, although Amazon Alexa did not perform as well on Hindi queries.  
We also found many occurrences where IPAs failed to process the emotion in the query and responded in a bland manner (by giving just web search results), sometimes even giving problematic and counter-productive responses. Our experiments offer evidence and point directions to augment EI displayed in the existing potential of IPAs.


\subsection{Sadness category}
\begin{figure}[h]%
    \centering
    \subfloat[\centering Comparison of the IPAs' performance for `Sadness' related queries in English]{{\includegraphics[width = 0.45\linewidth]{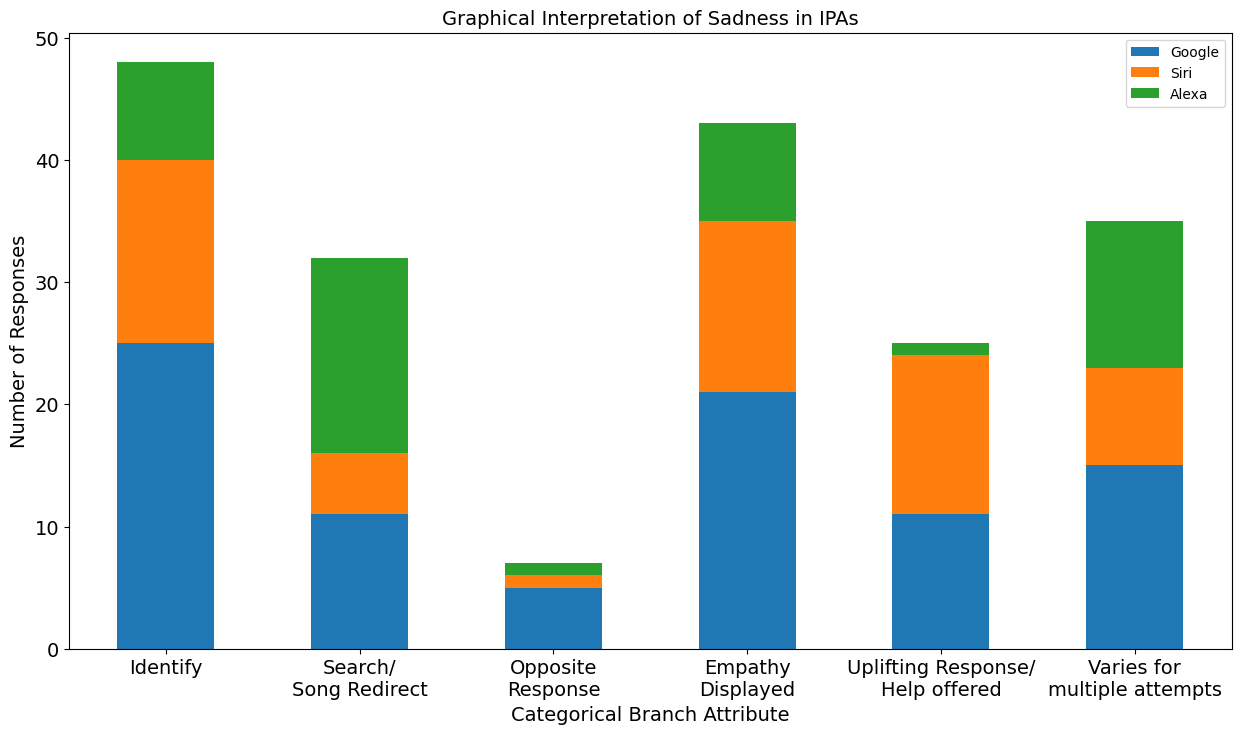} }}%
    \qquad
    \subfloat[\centering Variation of IPA performance across English and Hindi for `Sadness' related queries]{{\includegraphics[width = 0.45\linewidth]{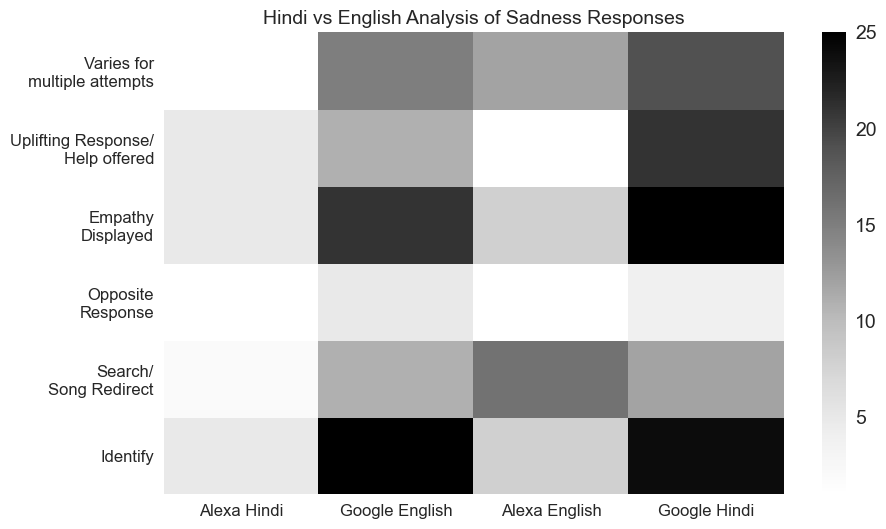} }}%
    \caption{\textbf{\textit{(a)}} In the inferences from the experiments conducted for EI for the `Sadness' emotion, Google Assistant was able to identify and show empathy to maximum number of queries posed. Amazon Alexa resorted to Search/Song redirects most and offerred the least uplifting responses. Apple Siri performed decently across the metrics and had the least search redirects compared to the remaining. \
    \textbf{\textit{(b)}} The heatplot for comparison between English and Hindi shows that Google Assistant's performance across the metrics was at par for both languages. In particular, its responses in Hindi were more empathetic that its English counterpart as well. Amazon Alexa's performance was relatively lower in Hindi compared to the same queries in English.}%
    \label{fig:sadness}%
\end{figure}

We found that Google Assistant outperformed the other 2 agents for sadness-related queries. The replies by Amazon Alexa were especially poor with it performing the maximum number of search redirects, and the least number of `helpful and uplifting' replies as can be seen by Fig \ref{fig:sadness} (a). The trend of Google Assistant giving more nuanced replies than Amazon Alexa remained across both languages of English and Hindi as can be seen by Fig \ref{fig:sadness} (b). Responses by Apple Siri were intermediary, though it performed the least number of search redirects. From Fig \ref{fig:sadness} (b) we see that Amazon Alexa's responses were better in the English language than Hindi, but both the languages were at par for Google Assistant, with Hindi replies having more points for the `uplifting / helpful' feature. This is a promising result pointing to the developing and evolving language support for Hindi. 
We find many problematic responses to `Sadness' related queries across all IPAs -- in these cases, the agents are not able to process the emotion correctly and give a detrimental response. For example, when we posed the query "I am no good!", the agent replied with "Got it!". In other cases, the IPAs are able to recognize the emotion and respond sympathetically. However, IPAs sometimes repeated the same responses across various queries, possibly choosing from a standard set of responses. For example, the response "I am sorry to hear that. Sometimes taking a quiet moment can help. You can try listening to your favourite music, or doing some simple stretches.", although displaying EI was repeated multiple times by Apple Siri across different queries.



\vspace{-2em}
\subsection{Humor category}

\begin{figure}[h]%
    \centering
    \subfloat[\centering Comparison of the IPAs' performance for `Humor' related queries in English]{{\includegraphics[width = 0.45\linewidth]{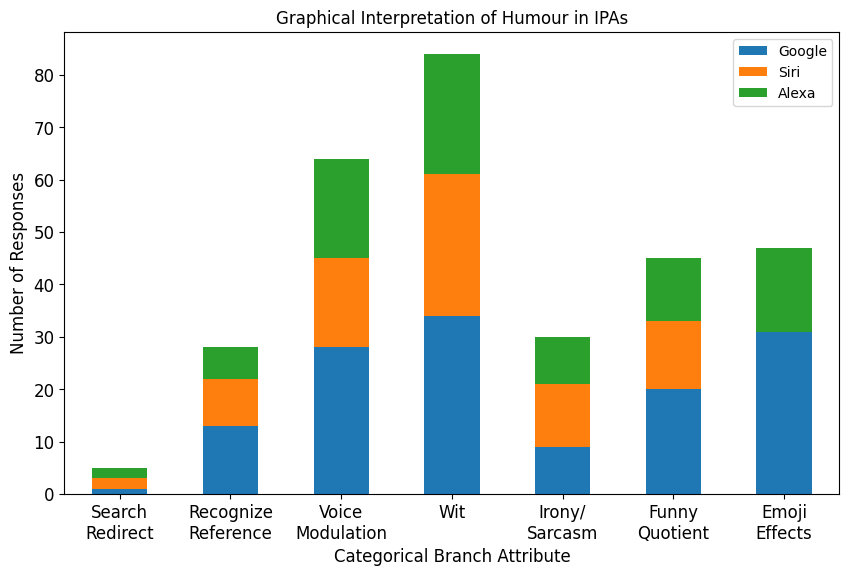} }}%
    \qquad
    \subfloat[\centering Variation of IPA performance across English and Hindi for `Humor' related queries]{{\includegraphics[width = 0.45\linewidth]{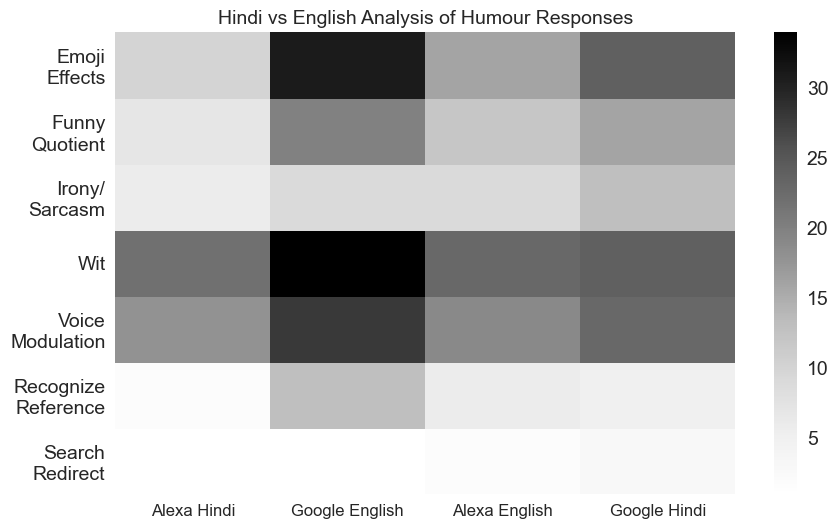} }}%
    \caption{\textbf{\textit{(a)}} In evalutating EI for the `Humour' emotion, all three IPAs have a significant fraction of witty responses. All IPAs also use voice modulation extensively. Siri does not utilize emojis in its conversation at all, however it has the largest proportion of ironic/sarcastic replies which are in turn quite human-like. The number of search redirects by all three virtual agents is drastically low compared to their responses in `Sadness'.     \
    \textbf{\textit{(b)}} As per the heatplot for comparison between English and Hindi, both Google Assistant and Amazon Alexa's performance matches in both languages to a good extent.}%
    \label{fig:humour}%
\end{figure}

We find IPAs performing similarly in the `Humor' category, with few fluctuations in ranking across different features as it can be observed in Fig \ref{fig:humour} (a). We find all agents use voice modulation to a large extent to convey humorous responses \ref{fig:humour} (a). We observe way fewer search redirects in the humor category than sadness (across all IPAs) indicating that agents are able to understand the query more accurately when they are tasked with giving a humorous response.  As can be seen in Fig \ref{fig:humour} (b), Google Assistant and Alexa's humorous responses in Hindi were not too far behind those in English. This is in contrast to Alexa's Hindi responses in the sadness category where it performed poorly. We also find that when comparing Google Assistant and Alexa's responses in Hindi, Fig \ref{fig:humour} (b), both agents were at par, unlike in the sadness field. It is also interesting to notice how, for some responses, both the agents utilize witty pop and Indian culture references in English and Hindi respectively. 

For example; in English, as a response to the query \textit{“What does the fox say”} (which is a popular song), Google assistant replies with, \textit{“Ring ding ding ding dingeringeding and wa pa pa pa pa pa pa pow. Or so I've heard”.} Similarly for Hindi, on asking Alexa, \textit{"Kya tum mujhse pyaar karti ho?"} , Alexa responds with \textit{"Dil cheez kya hai, aap meri jaan leejiye / Dil vil pyaar vyaar main kya jaanu re!"}, which are popular dialogues from Bollywood movie classics. 

While Amazon Alexa and Google Assistant perform well in the `wit' and `voice modulation` category, Google Assistant does have a slight edge over the others, with a higher fraction of  responses scored as 1 by the annotators.

Through the qualitative analysis of comments left by the annotators, we found the voice modulation feature was highly appreciated in the `Humor' category. Users were intrigued and claimed the feature improved the quality of the humorous response. Annotators believed emoji usage added more expression to a response.  Whilst Amazon Alexa and Google Assistant employed emojis to enhance their communication, Apple Siri did not.


\vspace{-1em}
\section{Discussion and Conclusion}

Our work is the first to probe the question of emotional intelligence (EI) of IPAs in Indian languages -- specifically Hindi. We study EI across 2 major categories -- `Sadness' and `Humor' -- by posing a myriad of queries and comparing the results of IPAs -- Amazon Alexa, Apple Siri, and Google Assistant.  We propose a new qualitative and quantitative evaluation scheme that builds on previous works and introduces new features that are useful to modern-day IPAs.

We find promising results for Hindi as Google Assistant returns appropriate and similar responses in both Hindi and English, signifying that efforts are being made to parse Hindi and respond to emotional queries. 
We also highlight cases where IPAs fail to understand an emotional query and respond in a bland or problematic manner. We find many such cases in our dataset, signifying that there is room for improvement in the EI displayed by IPAs. Finally, we make public the dataset of queries and responses by different IPAs that can be useful for further NLP and HCI research.

\bibliographystyle{ACM-Reference-Format}
\bibliography{sample-base}


\begin{thebibliography}{20}


\ifx \showCODEN    \undefined \def \showCODEN     #1{\unskip}     \fi
\ifx \showDOI      \undefined \def \showDOI       #1{#1}\fi
\ifx \showISBNx    \undefined \def \showISBNx     #1{\unskip}     \fi
\ifx \showISBNxiii \undefined \def \showISBNxiii  #1{\unskip}     \fi
\ifx \showISSN     \undefined \def \showISSN      #1{\unskip}     \fi
\ifx \showLCCN     \undefined \def \showLCCN      #1{\unskip}     \fi
\ifx \shownote     \undefined \def \shownote      #1{#1}          \fi
\ifx \showarticletitle \undefined \def \showarticletitle #1{#1}   \fi
\ifx \showURL      \undefined \def \showURL       {\relax}        \fi
\providecommand\bibfield[2]{#2}
\providecommand\bibinfo[2]{#2}
\providecommand\natexlab[1]{#1}
\providecommand\showeprint[2][]{arXiv:#2}

\bibitem[\protect\citeauthoryear{Abercrombie, Curry, Pandya, and
  Rieser}{Abercrombie et~al\mbox{.}}{2021}]%
        {va_gender_pronouns}
\bibfield{author}{\bibinfo{person}{Gavin Abercrombie},
  \bibinfo{person}{Amanda~Cercas Curry}, \bibinfo{person}{Mugdha Pandya}, {and}
  \bibinfo{person}{Verena Rieser}.} \bibinfo{year}{2021}\natexlab{}.
\newblock \showarticletitle{Alexa, Google, Siri: What are Your Pronouns? Gender
  and Anthropomorphism in the Design and Perception of Conversational
  Assistants}.
\newblock \bibinfo{journal}{\emph{CoRR}}  \bibinfo{volume}{abs/2106.02578}
  (\bibinfo{year}{2021}).
\newblock
\showeprint[arxiv]{2106.02578}
\urldef\tempurl%
\url{https://arxiv.org/abs/2106.02578}
\showURL{%
\tempurl}


\bibitem[\protect\citeauthoryear{Baki and Kim}{Baki and Kim}{2021}]%
        {google_nbu_voice}
\bibfield{author}{\bibinfo{person}{Asif Baki} {and} \bibinfo{person}{Jung~E
  Kim}.} \bibinfo{year}{2021}\natexlab{}.
\newblock \bibinfo{title}{How to help people navigate the internet,
  voice-first}.
\newblock
\newblock
\urldef\tempurl%
\url{https://blog.google/technology/next-billion-users/voice-users-playbook/}
\showURL{%
\tempurl}


\bibitem[\protect\citeauthoryear{Brackett and Salovey}{Brackett and
  Salovey}{2006}]%
        {msceit}
\bibfield{author}{\bibinfo{person}{Marc~A Brackett} {and}
  \bibinfo{person}{Peter Salovey}.} \bibinfo{year}{2006}\natexlab{}.
\newblock \showarticletitle{{Measuring emotional intelligence with the
  Mayer-Salovery-Caruso emotional intelligence test (MSCEIT)}}.
\newblock \bibinfo{journal}{\emph{Psicothema}}  \bibinfo{volume}{18}
  (\bibinfo{year}{2006}), \bibinfo{pages}{34--41}.
\newblock


\bibitem[\protect\citeauthoryear{Devaram}{Devaram}{2020}]%
        {empathetic_chatbots_mental_health}
\bibfield{author}{\bibinfo{person}{Sarada Devaram}.}
  \bibinfo{year}{2020}\natexlab{}.
\newblock \showarticletitle{Empathic Chatbot: Emotional Intelligence for
  Empathic Chatbot: Emotional Intelligence for Mental Health Well-being}.
\newblock \bibinfo{journal}{\emph{CoRR}}  \bibinfo{volume}{abs/2012.09130}
  (\bibinfo{year}{2020}).
\newblock
\showeprint[arxiv]{2012.09130}
\urldef\tempurl%
\url{https://arxiv.org/abs/2012.09130}
\showURL{%
\tempurl}


\bibitem[\protect\citeauthoryear{Ghafurian, Budnarain, and Hoey}{Ghafurian
  et~al\mbox{.}}{2019}]%
        {role_of_emotion}
\bibfield{author}{\bibinfo{person}{Moojan Ghafurian}, \bibinfo{person}{Neil
  Budnarain}, {and} \bibinfo{person}{Jesse Hoey}.}
  \bibinfo{year}{2019}\natexlab{}.
\newblock \showarticletitle{Role of emotions in perception of humanness of
  virtual agents}. In \bibinfo{booktitle}{\emph{Proceedings of the 18th
  International Conference on Autonomous Agents and MultiAgent Systems}}.
  \bibinfo{pages}{1979--1981}.
\newblock


\bibitem[\protect\citeauthoryear{Goleman}{Goleman}{1995}]%
        {ei_defn}
\bibfield{author}{\bibinfo{person}{Daniel. Goleman}.}
  \bibinfo{year}{1995}\natexlab{}.
\newblock \bibinfo{booktitle}{\emph{Emotional intelligence / Daniel Goleman}}.
\newblock \bibinfo{publisher}{Bantam Books New York}. xiv, 352 p. : pages.
\newblock
\showISBNx{055309503 0553375067 0747528306}


\bibitem[\protect\citeauthoryear{Hoy}{Hoy}{2018}]%
        {iva_def}
\bibfield{author}{\bibinfo{person}{Matthew Hoy}.}
  \bibinfo{year}{2018}\natexlab{}.
\newblock \showarticletitle{Alexa, Siri, Cortana, and More: An Introduction to
  Voice Assistants}.
\newblock \bibinfo{journal}{\emph{Medical Reference Services Quarterly}}
  \bibinfo{volume}{37} (\bibinfo{date}{01} \bibinfo{year}{2018}),
  \bibinfo{pages}{81--88}.
\newblock
\urldef\tempurl%
\url{https://doi.org/10.1080/02763869.2018.1404391}
\showDOI{\tempurl}


\bibitem[\protect\citeauthoryear{Hu, Huang, Hu, and Xu}{Hu
  et~al\mbox{.}}{2021}]%
        {acoustic_cues_voice}
\bibfield{author}{\bibinfo{person}{Jiaxiong Hu}, \bibinfo{person}{Yun Huang},
  \bibinfo{person}{Xiaozhu Hu}, {and} \bibinfo{person}{Yingqing Xu}.}
  \bibinfo{year}{2021}\natexlab{}.
\newblock \bibinfo{booktitle}{\emph{Enhancing the Perceived Emotional
  Intelligence of Conversational Agents through Acoustic Cues}}.
\newblock \bibinfo{publisher}{Association for Computing Machinery},
  \bibinfo{address}{New York, NY, USA}.
\newblock
\showISBNx{9781450380959}
\urldef\tempurl%
\url{https://doi.org/10.1145/3411763.3451660}
\showURL{%
\tempurl}


\bibitem[\protect\citeauthoryear{Lopatovska}{Lopatovska}{2020}]%
        {humour_classification}
\bibfield{author}{\bibinfo{person}{Irene Lopatovska}.}
  \bibinfo{year}{2020}\natexlab{}.
\newblock \showarticletitle{Classification of humorous interactions with
  intelligent personal assistants}.
\newblock \bibinfo{journal}{\emph{Journal of Librarianship and Information
  Science}} \bibinfo{volume}{52}, \bibinfo{number}{3} (\bibinfo{year}{2020}),
  \bibinfo{pages}{931--942}.
\newblock
\urldef\tempurl%
\url{https://doi.org/10.1177/0961000619891771}
\showDOI{\tempurl}
\showeprint{https://doi.org/10.1177/0961000619891771}


\bibitem[\protect\citeauthoryear{Lopatovska and Williams}{Lopatovska and
  Williams}{2018}]%
        {personification_alexa}
\bibfield{author}{\bibinfo{person}{Irene Lopatovska} {and}
  \bibinfo{person}{Harriet Williams}.} \bibinfo{year}{2018}\natexlab{}.
\newblock \showarticletitle{Personification of the Amazon Alexa: BFF or a
  Mindless Companion}. In \bibinfo{booktitle}{\emph{Proceedings of the 2018
  Conference on Human Information Interaction \& Retrieval}} (New Brunswick,
  NJ, USA) \emph{(\bibinfo{series}{CHIIR '18})}.
  \bibinfo{publisher}{Association for Computing Machinery},
  \bibinfo{address}{New York, NY, USA}, \bibinfo{pages}{265–268}.
\newblock
\showISBNx{9781450349253}
\urldef\tempurl%
\url{https://doi.org/10.1145/3176349.3176868}
\showDOI{\tempurl}


\bibitem[\protect\citeauthoryear{L{\'o}pez, Quesada, and Guerrero}{L{\'o}pez
  et~al\mbox{.}}{2018}]%
        {speech_voice_NUI}
\bibfield{author}{\bibinfo{person}{Gustavo L{\'o}pez}, \bibinfo{person}{Luis
  Quesada}, {and} \bibinfo{person}{Luis~A. Guerrero}.}
  \bibinfo{year}{2018}\natexlab{}.
\newblock \showarticletitle{Alexa vs. Siri vs. Cortana vs. Google Assistant: A
  Comparison of Speech-Based Natural User Interfaces}. In
  \bibinfo{booktitle}{\emph{Advances in Human Factors and Systems
  Interaction}}, \bibfield{editor}{\bibinfo{person}{Isabel~L. Nunes}} (Ed.).
  \bibinfo{publisher}{Springer International Publishing},
  \bibinfo{address}{Cham}, \bibinfo{pages}{241--250}.
\newblock
\showISBNx{978-3-319-60366-7}


\bibitem[\protect\citeauthoryear{Ma, Yang, and Fung}{Ma et~al\mbox{.}}{2019}]%
        {personality_perceived_ei}
\bibfield{author}{\bibinfo{person}{Xiaojuan Ma}, \bibinfo{person}{Emily Yang},
  {and} \bibinfo{person}{Pascale Fung}.} \bibinfo{year}{2019}\natexlab{}.
\newblock \showarticletitle{Exploring Perceived Emotional Intelligence of
  Personality-Driven Virtual Agents in Handling User Challenges}. In
  \bibinfo{booktitle}{\emph{The World Wide Web Conference}} (San Francisco, CA,
  USA) \emph{(\bibinfo{series}{WWW '19})}. \bibinfo{publisher}{Association for
  Computing Machinery}, \bibinfo{address}{New York, NY, USA},
  \bibinfo{pages}{1222–1233}.
\newblock
\showISBNx{9781450366748}
\urldef\tempurl%
\url{https://doi.org/10.1145/3308558.3313400}
\showDOI{\tempurl}


\bibitem[\protect\citeauthoryear{Miner, Milstein, Schueller, Hegde, Mangurian,
  and Linos}{Miner et~al\mbox{.}}{2016}]%
        {jamainternmed}
\bibfield{author}{\bibinfo{person}{Adam~S. Miner}, \bibinfo{person}{Arnold
  Milstein}, \bibinfo{person}{Stephen Schueller}, \bibinfo{person}{Roshini
  Hegde}, \bibinfo{person}{Christina Mangurian}, {and} \bibinfo{person}{Eleni
  Linos}.} \bibinfo{year}{2016}\natexlab{}.
\newblock \showarticletitle{{Smartphone-Based Conversational Agents and
  Responses to Questions About Mental Health, Interpersonal Violence, and
  Physical Health}}.
\newblock \bibinfo{journal}{\emph{JAMA Internal Medicine}}
  \bibinfo{volume}{176}, \bibinfo{number}{5} (\bibinfo{date}{05}
  \bibinfo{year}{2016}), \bibinfo{pages}{619--625}.
\newblock
\showISSN{2168-6106}
\urldef\tempurl%
\url{https://doi.org/10.1001/jamainternmed.2016.0400}
\showDOI{\tempurl}
\showeprint{https://jamanetwork.com/journals/jamainternalmedicine/articlepdf/2500043/ioi160007.pdf}


\bibitem[\protect\citeauthoryear{Reis, Paulino, Paredes, and Barroso}{Reis
  et~al\mbox{.}}{2017}]%
        {elderly_social_bond}
\bibfield{author}{\bibinfo{person}{Ars{\'e}nio Reis}, \bibinfo{person}{Dennis
  Paulino}, \bibinfo{person}{Hugo Paredes}, {and} \bibinfo{person}{Jo{\~a}o
  Barroso}.} \bibinfo{year}{2017}\natexlab{}.
\newblock \showarticletitle{Using Intelligent Personal Assistants to Strengthen
  the Elderlies' Social Bonds}. In \bibinfo{booktitle}{\emph{Universal Access
  in Human--Computer Interaction. Human and Technological Environments}},
  \bibfield{editor}{\bibinfo{person}{Margherita Antona} {and}
  \bibinfo{person}{Constantine Stephanidis}} (Eds.).
  \bibinfo{publisher}{Springer International Publishing},
  \bibinfo{address}{Cham}, \bibinfo{pages}{593--602}.
\newblock
\showISBNx{978-3-319-58700-4}


\bibitem[\protect\citeauthoryear{Schuller and Schuller}{Schuller and
  Schuller}{2018}]%
        {age_artificial_ei}
\bibfield{author}{\bibinfo{person}{Dagmar Schuller} {and}
  \bibinfo{person}{Bj\"{o}rn~W. Schuller}.} \bibinfo{year}{2018}\natexlab{}.
\newblock \showarticletitle{The Age of Artificial Emotional Intelligence}.
\newblock \bibinfo{journal}{\emph{Computer}} \bibinfo{volume}{51},
  \bibinfo{number}{9} (\bibinfo{date}{Sept.} \bibinfo{year}{2018}),
  \bibinfo{pages}{38–46}.
\newblock
\showISSN{0018-9162}
\urldef\tempurl%
\url{https://doi.org/10.1109/MC.2018.3620963}
\showDOI{\tempurl}


\bibitem[\protect\citeauthoryear{Tulshan and Dhage}{Tulshan and Dhage}{2019}]%
        {iva_survey}
\bibfield{author}{\bibinfo{person}{Amrita~S. Tulshan} {and}
  \bibinfo{person}{Sudhir~Namdeorao Dhage}.} \bibinfo{year}{2019}\natexlab{}.
\newblock \showarticletitle{Survey on Virtual Assistant: Google Assistant,
  Siri, Cortana, Alexa}. In \bibinfo{booktitle}{\emph{Advances in Signal
  Processing and Intelligent Recognition Systems}},
  \bibfield{editor}{\bibinfo{person}{Sabu~M. Thampi}, \bibinfo{person}{Oge
  Marques}, \bibinfo{person}{Sri Krishnan}, \bibinfo{person}{Kuan-Ching Li},
  \bibinfo{person}{Domenico Ciuonzo}, {and} \bibinfo{person}{Maheshkumar~H.
  Kolekar}} (Eds.). \bibinfo{publisher}{Springer Singapore},
  \bibinfo{address}{Singapore}, \bibinfo{pages}{190--201}.
\newblock


\bibitem[\protect\citeauthoryear{Waytz, Heafner, and Epley}{Waytz
  et~al\mbox{.}}{2014}]%
        {autonomous_vehicles_mind}
\bibfield{author}{\bibinfo{person}{Adam Waytz}, \bibinfo{person}{Joy Heafner},
  {and} \bibinfo{person}{Nicholas Epley}.} \bibinfo{year}{2014}\natexlab{}.
\newblock \showarticletitle{The mind in the machine: Anthropomorphism increases
  trust in an autonomous vehicle}.
\newblock \bibinfo{journal}{\emph{Journal of Experimental Social Psychology}}
  \bibinfo{volume}{52} (\bibinfo{year}{2014}), \bibinfo{pages}{113--117}.
\newblock
\showISSN{0022-1031}
\urldef\tempurl%
\url{https://doi.org/10.1016/j.jesp.2014.01.005}
\showDOI{\tempurl}


\bibitem[\protect\citeauthoryear{Yang, Lavie, Dyer, and Hovy}{Yang
  et~al\mbox{.}}{2015}]%
        {latent-humour}
\bibfield{author}{\bibinfo{person}{Diyi Yang}, \bibinfo{person}{Alon Lavie},
  \bibinfo{person}{Chris Dyer}, {and} \bibinfo{person}{Eduard Hovy}.}
  \bibinfo{year}{2015}\natexlab{}.
\newblock \showarticletitle{Humor Recognition and Humor Anchor Extraction}. In
  \bibinfo{booktitle}{\emph{Proceedings of the 2015 Conference on Empirical
  Methods in Natural Language Processing}}. \bibinfo{publisher}{Association for
  Computational Linguistics}, \bibinfo{address}{Lisbon, Portugal},
  \bibinfo{pages}{2367--2376}.
\newblock
\urldef\tempurl%
\url{https://doi.org/10.18653/v1/D15-1284}
\showDOI{\tempurl}


\bibitem[\protect\citeauthoryear{Yang, Ma, and Fung}{Yang
  et~al\mbox{.}}{2017a}]%
        {ei_ipa}
\bibfield{author}{\bibinfo{person}{Yang Yang}, \bibinfo{person}{Xiaojuan Ma},
  {and} \bibinfo{person}{Pascale Fung}.} \bibinfo{year}{2017}\natexlab{a}.
\newblock \showarticletitle{Perceived Emotional Intelligence in Virtual
  Agents}. In \bibinfo{booktitle}{\emph{Proceedings of the 2017 CHI Conference
  Extended Abstracts on Human Factors in Computing Systems}} (Denver, Colorado,
  USA) \emph{(\bibinfo{series}{CHI EA '17})}. \bibinfo{publisher}{Association
  for Computing Machinery}, \bibinfo{address}{New York, NY, USA},
  \bibinfo{pages}{2255–2262}.
\newblock
\showISBNx{9781450346566}
\urldef\tempurl%
\url{https://doi.org/10.1145/3027063.3053163}
\showDOI{\tempurl}


\bibitem[\protect\citeauthoryear{Yang, Ma, and Fung}{Yang
  et~al\mbox{.}}{2017b}]%
        {perceived_ei}
\bibfield{author}{\bibinfo{person}{Yang Yang}, \bibinfo{person}{Xiaojuan Ma},
  {and} \bibinfo{person}{Pascale Fung}.} \bibinfo{year}{2017}\natexlab{b}.
\newblock \showarticletitle{Perceived Emotional Intelligence in Virtual
  Agents}. In \bibinfo{booktitle}{\emph{Proceedings of the 2017 CHI Conference
  Extended Abstracts on Human Factors in Computing Systems}} (Denver, Colorado,
  USA) \emph{(\bibinfo{series}{CHI EA '17})}. \bibinfo{publisher}{Association
  for Computing Machinery}, \bibinfo{address}{New York, NY, USA},
  \bibinfo{pages}{2255–2262}.
\newblock
\showISBNx{9781450346566}
\urldef\tempurl%
\url{https://doi.org/10.1145/3027063.3053163}
\showDOI{\tempurl}


\end{thebibliography}










\end{document}